\documentclass[
aip,
reprint,
superscriptaddress,
longbibliography,
showkeys,
 amsmath,amssymb,
 aps,
pre, 
]{revtex4-2}

\usepackage{graphicx}
\usepackage[caption=false]{subfig} 
\usepackage{changes}

\usepackage{float}
     
\usepackage{dcolumn}
\usepackage{bm}
\usepackage{geometry}
\newcommand{\ee}{\end{equation}} 
\newcommand{\be}{\begin{equation}}

\usepackage[mathscr]{euscript}  
\usepackage{xcolor}  

\usepackage{tcolorbox}
\usepackage{comment}
\usepackage{float}
\def\la{\langle}

\def\ra{\rangle}

\makeatletter
\newsavebox{\@brx}
\newcommand{\llangle}[1][]{\savebox{\@brx}{\(\m@th{#1\langle}\)}%
  \mathopen{\copy\@brx\kern-0.5\wd\@brx\usebox{\@brx}}}
\newcommand{\rrangle}[1][]{\savebox{\@brx}{\(\m@th{#1\rangle}\)}%
  \mathclose{\copy\@brx\kern-0.5\wd\@brx\usebox{\@brx}}}
\makeatother

\begin{document} 

\preprint{APS/123-QED}

\title{Brownian motion with stochastic energy renewals}

\author{Ion Santra}
\thanks{\textit{Correspondence: ion.santra@theorie.physik.uni-goettingen.de}}
\affiliation{Institute for Theoretical Physics, University of G\"{o}ttingen, 37077 G\"{o}ttingen,Germany}

\author{Kristian St\o{}levik Olsen}
\thanks{\textit{Correspondence: kristian.olsen@hhu.de}}
\affiliation{Institut für Theoretische Physik II - Weiche Materie, Heinrich-Heine-Universität Düsseldorf, D-40225 Düsseldorf, Germany}

\begin{abstract}
We investigate the impact of intermittent energy injections on a Brownian particle, modeled as stochastic renewals of its kinetic energy to a fixed value. Between renewals, the particle follows standard underdamped Langevin dynamics. For energy renewals occurring at a constant rate, we find non-Boltzmannian energy distributions that undergo a shape transition driven by the competition between the velocity relaxation timescale and the renewal timescale. In the limit of rapid renewals, the dynamics mimics one-dimensional run-and-tumble motion, while at finite renewal rates, the effective diffusion coefficient exhibits non-monotonic behavior. To quantify the system's departure from equilibrium, we derive a modified fluctuation-response relation and demonstrate the absence of a consistent effective temperature. The dissipation is characterized by deviations from equilibrium-like response, captured via the Harada–Sasa relation. Finally, we extend the analysis to non-Poissonian renewal processes and introduce a dimensionless conversion coefficient that quantifies the thermodynamic cost of diffusion.
\end{abstract}

\maketitle

\section{Introduction}

{\color{black}{


For systems close to equilibrium, results from classical statistical mechanics impose strong constraints on the dynamics and energetics \cite{kubo1966fluctuation,kubo2012statistical,risken1996fokker}. A particularly intriguing class of systems driven from equilibrium are those that repeatedly receive non-thermal energetic input, through for example feeding, collisions or other environmental interactions, leading to a breakdown of detailed balance and the equipartition of energy ~\cite{ramaswamy2010mechanics,marchetti2013hydrodynamics,mauri2006violation,bellon2001violation,abou2004probing}.

Random energy injections can arise in a variety of contexts [See Fig.~\ref{fig:sketch}]. In active matter, for example, living organisms or synthetic particles extract energy from their environment to propel themselves, often taking place through mechanisms such as chemical reactions~\cite{ke2010motion,ebbens2011direct} or light absorption~\cite{palacci2014light,vutukuri2020light}. Similar dynamics can emerge in non-active matter contexts: a tracer particle in a dilute bath where rare but strong interactions lead to kinetic energy increments \cite{park2020rapid,leptos2009dynamics}, or a heterogeneous environment with hot-spots that boost the particle's energy. Another important example is found in driven granular gases, where particles lose energy through inelastic collisions but receive intermittent energy kicks from external driving or boundary vibrations, leading to non-Boltzmann kinetic energy distributions~\cite{rouyer2000velocity,van2004velocity}. The common feature is a direct competition between the steady thermal relaxation and non-thermal stochastic energy input. 

To investigate such dynamics in a minimal setting, we consider an inertial Brownian particle whose kinetic energy is intermittently reset to a fixed value, modeling random energy injections. Between energy boosts/renewals, the particle undergoes standard Langevin dynamics, leading to energy dissipation and relaxation toward thermal equilibrium.
Our model shares features with established active matter descriptions, such as particles with internal energy depots ~\cite{schweitzer1998complex,erdmann2000brownian,ebeling1999active,schweitzer2003brownian,zhang2008active,miranda2025collective,olsen2021active,bienewald2025hungrydaemonenergyharvestingactive}. However, rather than explicitly modeling both energy absorption and conversion into mechanical energy as continuous processes, here we present a simpler and analytically tractable model with discrete energy injections, offering insights into energetics, dynamics, and their interplay.

\begin{figure}[t!]
    \centering
    \includegraphics[width=8.6cm]{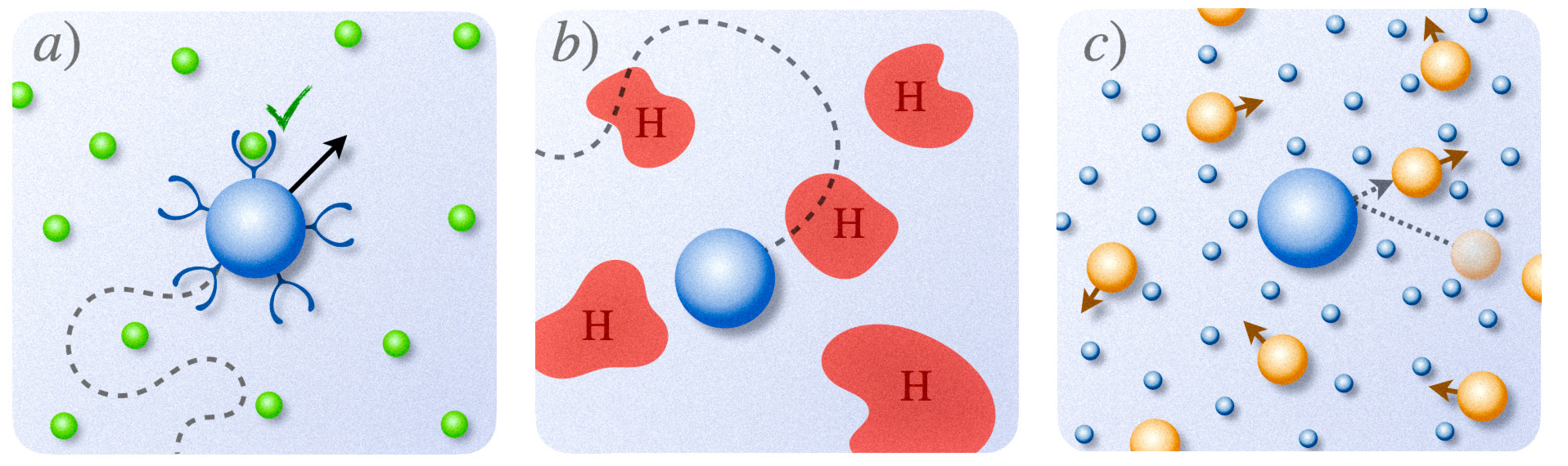}
    \caption{a) A particle moving through an environment with scarce resources intermittently gain energy upon resource harvesting. b) Heterogeneous media with hot-spots may randomly boost the particle motion. c) Energetic boosts may occur for tracers in dilute active baths. }
    \label{fig:sketch}
\end{figure}

We begin by analyzing the case where energy renewals occur at a constant (Poissonian) rate. We compute the mean entropy production and energy dissipation, and show that higher-order energy fluctuations exhibit a non-monotonic dependence on the renewal rate. The stationary energy distribution is distinctly non-Boltzmannian, displaying a shape transition and exponential tails modified by power-law corrections. We then investigate the influence of energy renewals on spatial dynamics by computing the position moments. In the regime where the renewal timescale is shorter than the velocity relaxation timescale, the model exactly reproduces that of a run-and-tumble particle. More generally, the resulting dynamics shares features with particles driven by shot noise \cite{di2024brownian,fiasconaro2013controlling,fiasconaro2009tuning,strefler2009dynamics} or intermittent active motion \cite{datta2024random,santra2024dynamics,olsen2024optimal}. We further study the linear response of such a particle to an external perturbation, and leading to a modified fluctuation response relation, and also revealing the lack of a consistent effective temperature description. We show that the deviation from the equilibrium-like response can be interpreted as a measure of dissipation, quantified using the well-known Harada–Sasa relation~\cite{harada2005equality,harada2006energy}. Finally, we extend our analysis to non-Poissonian energy renewals and relate the spatial dynamics to energetics by computing a conversion coefficient that provides a measure of the thermodynamic cost of diffusion.

The remainder of our paper is organized as follows. We introduce the model in Sec.~\ref{sec:model}, then we explore the stationary energy fluctuations for the Poisson case in~Sec.~\ref{ssec:energyfl}. Thereafter, we study the impact of energy renewals on spatial dynamics, and linear response in Sec.~\ref{sec:spacex}. General waiting-times for energy renewals are studied in Sec.~\ref{ssec:thermo}. Finally, a concluding discussion is offered in Sec.~\ref{sec:concl}.
}}

\section{Model and Energetics}\label{sec:model}

For a freely moving inertial Brownian particle, the equation of motion is given by the Langevin equations
\begin{align}
    \dot x(t) &= v(t),\\
    m\dot v(t) &= - \gamma v(t) + \eta(t),
    \label{eq:underdamped}
\end{align}
where $\eta(t)$ is a Gaussian white noise with zero mean and autocorrelation $\langle\eta(t)\eta(0)\rangle= 2 k_BT \gamma\delta(t)$.
The dynamics of the kinetic energy $K=mv^2/2$, conditioned on a specific initial velocity, is
\begin{equation} \label{eq:kin0}
     K^{(0)}(t|K_0^{(0)}) = K_0^{(0)} e^{- 2 \gamma t/m} + \frac{k_BT}{2} (1-e^{-2\gamma t/m}) ,
\end{equation}
where at late times equipartition is reached. Intermittent energy consumption, modeled as energy resetting events, drives the system away from thermal equilibrium. At random times $\{t_j\}$ the particle energy is reset
\begin{equation}
    K(t_j) \to K_R. 
\end{equation}
Such random energy renewals break the equipartition and fluctuation-dissipation theorems, resulting in non-trivial energetics.

\subsection{The first law and dissipation rate}
\begin{figure}
    \centering
    \includegraphics[width=8.6cm]{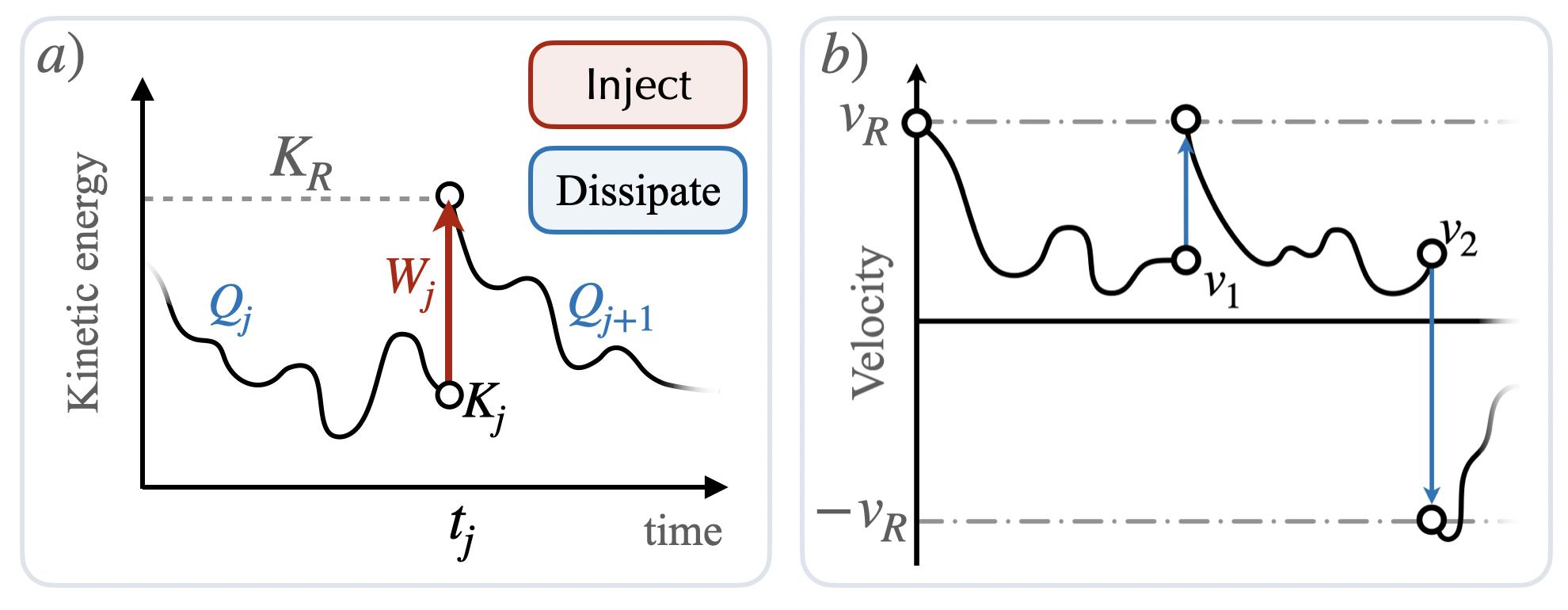}
    \caption{a) Kinetic energy as a function of time, with dissipation and work input, i.e. energy injections. b) The simplest velocity rule consistent with energy resets is an unbiased one where velocity resets to $\pm v_R$ with equal probability.}
    \label{fig:resets}
\end{figure}

 Deviations from thermal equilibrium can be measured by the dissipation rate, or entropy production, which in this case will be non-zero due to the sustained energy injections. Assume that the particle starts with energy $K_0$, and let $t_j$ denote the times at which energy is renewed to the value $K_R$. We let there be $n$ such renewal events in a total time $t$, and let $K(t)$ be the final energy. The energy at the instant just before the $j$th renewal we denote $K_j \equiv K(t_j^-)$.  The dissipated energy $Q_j$ during a time-interval $(t_{j-1},t_j)$, $j = 2,...,n$ is simply $Q_j = K_R - K_j$, while the energy dissipated during the final incomplete renewal period is then $Q_{n+1} = K_R - K(t)$. See Fig.~\ref{fig:resets} (a) for a sketch of the dynamics. For $j\geq 2$, the dissipated energy is exactly the work $W_j$ injected at each instant $t_j$. Hence, the model by construction satisfies
\begin{equation}
    W - Q = K(t) - K_0 \equiv \Delta K
\end{equation}
where $W = \sum_{j=1}^n W_j$ and $Q = \sum_{j=1}^{n+1} Q_j$ are the total work done to the system, and heat dissipated into the bath respectively. This is a version of the first law of thermodynamics, valid not only on average but for each stochastic realization.

The mean dissipation can be calculated as 
\begin{equation}
   \langle Q\rangle = \sum_{j=1}^{n+1} \langle Q_j\rangle  = K_0 -  \langle K(t)\rangle + n(K_R - \langle K(\tau)\rangle)
\end{equation}
where we used the fact that each $K_j$ is an independent random variables with $\langle K_j\rangle = \langle K(\tau)\rangle$, where both thermal noise and duration $\tau$ are averaged over.  At late times, or equivalently large $n$, we can ignore the bounded terms $K_0 - \langle K(t)\rangle$. Using the fact that $n =  rt$, one readily finds the rate of dissipation
\begin{equation}
     \dot Q \equiv  \lim_{t\to \infty}\frac{\langle Q\rangle}{t} = r (K_R - \langle K(\tau)\rangle)
\end{equation}
Here $K(\tau)$ is easily calculated from the free (non-renewing) dynamics, Eq. (\ref{eq:kin0}), resulting in 
\begin{equation}\label{eq:pi}
    \dot Q = \frac{2 r}{2 + \frac{r m}{ \gamma}}\left(K_R - \frac{k_BT}{2} \right)
\end{equation}

Several interesting observations are worth mentioning. First, being proportional to the rate $r$, we see that energy renewals are the main cause of departure from thermal equilibrium. Secondly, in appropriate limits the above  dissipation rate takes recognizable forms. For large rates $r\gg \gamma/m$ and $K_R \gg k_BT$, we simply have $\dot Q = \gamma v_R^2$, where we used $K_R = \frac{1}{2} m v_R^2$. This is the dissipation rate for a particle moving at speed $v_R$ through a Stokesian medium with friction $\gamma$. When the reset rate is small, we find the leading order behavior $\dot Q\approx  r (K_R -\frac{k_BT}{2})$. In this limit, the energy given to the system on top of the thermal energy is allowed to completely dissipate before another reset takes place. It is also worth noting that while we mainly will consider the case where $K_R$ is greater than the typical thermal energies, the dissipation could also be negative if each renewal slows the particle down. This could be the case for example during collisions with an environment where energy is lost.

Eq. (\ref{eq:pi}) also shares strong similarities to the entropy production rate of inertial active particles \cite{frydel2023entropy}. This should not come as a surprise, since by constantly fueling the particle with energy, active motion is expected to emerge. In later sections we will explicitly show how the spatial dynamics of particles with energy resets exactly reproduces that of active particles in the correct limit. 

Another curious property of the above dissipation rate is that when energy is renewed to a value consistent with the equipartition theorem, $K_R = \frac{k_BT}{2} $, the (mean) dissipation rate vanishes and is not sufficient to detect any nonequilibrium effects even through the renewals strongly affect overall system behavior. To detect nonequilibrium effects in this case, one needs to study higher-order fluctuations, or consider the full distribution of energies, which we consider next.

\subsection{Energy fluctuations}\label{ssec:energyfl}
To gain further insights into the energetics of the system, we study here the full statistics of the particle's kinetic energy. We begin by deriving a hierarchy for the energetic moments, before proceeding to an exact derivation for the steady state energy distribution, which shows intriguing nonequilibrium behaviors. From here onwards, we renew the energy to its initial value by setting $K_R = K_0$. 

\subsubsection{Stationary Moments}
The kinetic energy of the particle without energy resetting follows a Langevin equation, which using Eq.~\eqref{eq:underdamped}, can be written as,
\begin{align}
    \dot{K}(t)=-\frac{2\gamma}{m} K(t)+\sqrt{\frac{2\gamma}{\beta m}K(t)}\,\eta(t).\label{eq:KE-langevin}
\end{align}
When $K(t)$ is reset to $K_0$ at a rate $r$, the distribution of energy evolves by the Fokker-Planck equation\footnote{we use a Stratonovich interpretation of the multiplicative noise in Eq.~\eqref{eq:KE-langevin}},
\begin{align}
    \partial_t P(K) &=\frac{2\gamma}{\beta m} \partial_K\left[ \beta KP(K)+\frac {\gamma}{2\beta m}P(K) \right.\cr&\left.+\frac{\gamma}{\beta m}K\partial_KP(K)
 \right]-rP(K)+r\delta(K-K_0).
 \label{eq:FP_KE}
\end{align}
The last two terms in the above equation account for the Poissonian renewal of energy to $K_0$ at rate $r$. Note that, Eq.~\eqref{eq:KE-langevin} involves a multiplicative noise, and we have used the Stratonovich prescription to write Eq.~\eqref{eq:FP_KE}.

To quantify the fluctuations of $K$ in the presence of the energy renewals, we analyze the stationary moments of the kinetic energy distribution. Multiplying the stationary Fokker-Planck equation,  Eq.~\eqref{eq:FP_KE} with $\partial_tP(K)=0$, by $K^n$ and integrating over $K$, we get a recursion equation for the $n$th order moment $M_n=\displaystyle\int dK\,K^n P(K)$,
\begin{align}
    \left(2n\frac{\gamma}{m}+r\right)M_n=\frac{2\gamma}{\beta m} \gamma \left(n^2-\frac n2\right)M_{n-1}+rK_0^n.
\end{align}
The above recursion relation can be exactly solved with the initial condition $M_0=1$, leading to the general form for the $n$th moment,
\begin{align}
   \beta^n M_n&=\frac{mr\Gamma (2 n+1)}{ 2^{2 n+1}\gamma}  \left(\frac{\Gamma \left(\frac{m r}{2 \gamma }\right) \, _1F_1\left(\frac{m r}{2 \gamma };\frac{1}{2};\alpha \right)}{\Gamma \left(n+\frac{m r}{2 \gamma }+1\right)}\right.\nonumber\\
    &\left.-\sqrt{\pi } \alpha ^{n+1} \, _2\tilde{F}_2\left(1,n+\frac{m r}{2 \gamma }+1;n+\frac{3}{2},n+2;\alpha \right)\right),
\end{align}
where we have chosen $K_0=\frac 12m v_0^2$ and defined
\begin{align}
    \alpha = \frac{K_0}{k_BT}
\end{align}
as a dimensionless measure of the renewed energy. From the general expression, the mean stationary kinetic energy is
\begin{align}
    M_1=\beta^{-1}\frac{\gamma +\alpha  m r}{2 \gamma +m r},
\end{align}
which corresponds to the equipartition of energy for $r=0$, and saturates to $\alpha$ for very frequent energy renewals $r\to\infty$. For $\alpha=1/2$, the average stationary kinetic energy is independent of the renewal rate, and takes the value $M_1 = k_BT/2$, consistent with energy equipartition. 

\begin{figure}
    \centering
    \includegraphics[width=\linewidth]{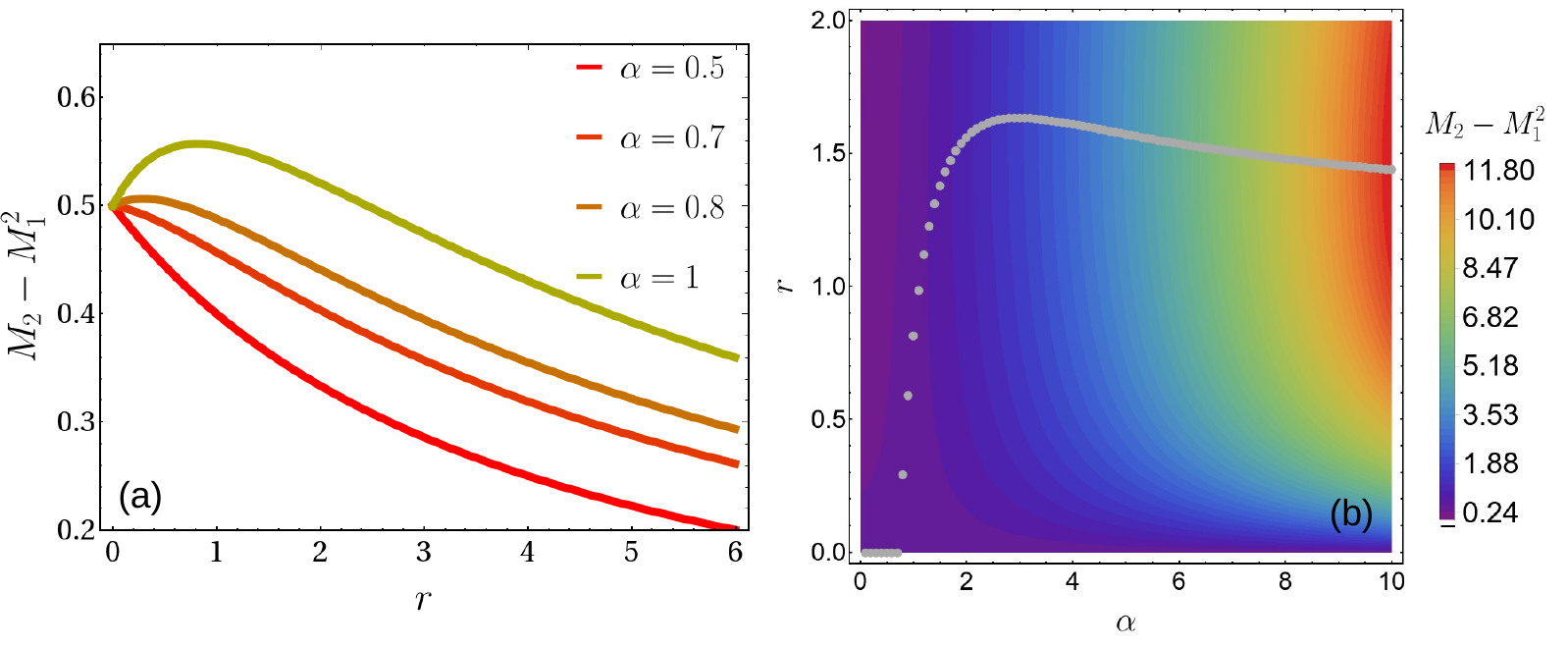}
    \caption{Kinetic energy fluctuations: (a) Variance of the kinetic energy for $m=\gamma=1$, and different values of $\alpha$, as given in Eq.~\eqref{ke:var}. (b) Kinetic energy variance in the $(r,\alpha)$ plane for same parameters as  (a). The dotted grey line denotes the optimal resetting rate at which the fluctuations are maximized.}
    \label{f:KE-var}
\end{figure}

The fluctuations of the kinetic energy about the mean is characterized by its variance,
\begin{align}
    M_2-M_1^2=\frac{6 \gamma ^2+\alpha ^2 m^2 r^2+2 (\alpha +3) \alpha  \gamma  m r}{\beta^2(2 \gamma +m r) (4 \gamma +m r)}.\label{ke:var}
\end{align}
The limiting value of the fluctuations for small renewal rate $r\to 0$ is given by $(k_BT)^2/2$ ; on the other hand, in the limit of $r\to \infty$ it goes to zero since in this limit the kinetic energy almost remains fixed at $K_0$. The more interesting limit is for $\alpha=1/2$, where the energy renewals are consistent with the equipartition of energy. In this case, the variance reduces to $\frac{2 \gamma (k_BT)^2 }{4 \gamma +m r}$; the explicit dependence on $r$ is the signature of the nonequilibrium nature of the dynamics. In this case, for any finite $r$ the energy fluctuations are decreased due to the renewals.

In general, the fluctuations exhibit a non-monotonic dependence on the energy renewal rate for $\alpha > \alpha_c = \left(\sqrt{6}-1\right)/2 \approx 0.725$: the variance increases with $r$ up to a certain value $r^*$, beyond which it decreases. In contrast, for smaller values $0 < \alpha < \alpha_c$, the fluctuations decay monotonically with increasing $r$. This behavior can be understood in terms of the typical region of the kinetic energy distribution in the absence of resetting, given by $\frac{\sqrt{\beta }\, e^{-\beta K}}{\sqrt{\pi K}}$. For small $\alpha$, the energy renewals predominantly occur within this typical region, resulting in a suppression of fluctuations. However, when $\alpha$ exceeds a critical threshold, the renewals are more likely to inject energy values far from the typical range, initially enhancing fluctuations until frequent renewal eventually dominates and suppresses the fluctuations again.

\subsubsection{Energy distribution}
Finally, we consider the full distribution of the particle's energy. This takes the form
\begin{equation}
     P(K) = \int dv \mathcal{P}(v) \delta(K - \frac{1}{2}m v^2)
\end{equation}
where we by $ \mathcal{P}(v)$ denote the steady-state velocity distribution of the particle under renewals.  Expanding the Dirac delta-function we arrive at
\begin{equation}\label{eq:ss_ke}
    {P}(K) = \frac{\mathcal{P}(\sqrt{2K/m})+\mathcal{P}(-\sqrt{2K/m})}{\sqrt{2 K/m}}
\end{equation}
given in terms of the steady state velocity distribution $\mathcal{P}(v)$. This follows from well-established results in resetting literature. Indeed, the energy resetting rule can be seen to result from the velocity resetting rule
\begin{equation}
    v(t_n)\to \pm v_0
    \label{velocity-reset}
\end{equation}
where positive and negative signs are taken with equal probability [see Fig.~\ref{fig:resets} (b)]. This can be encoded in a resetting distribution 
\begin{equation}
    \mathscr{P}_R(v) = \frac{\delta(v-v_0) + \delta(v+v_0)}{2}
    \label{reset-dist}
\end{equation}
so that at each reset, a new velocity is drawn from $\mathscr{P}_R(v)$.  The steady state of such processes takes the form \cite{evans2020stochastic}
\begin{equation}\label{eq:withresetting}
    \mathcal{P}(v) = r \int dv_0 \mathscr{P}_R(v_0) \tilde{\mathcal{P}}_0(v,r| v_0)
\end{equation} 
where tildes denote Laplace transforms, and the subscript $0$ denotes distributions without resetting. In this expression $\tilde{\mathcal{P}}_0(v,r| v_0)$ is simply the Laplace transform of the Ornstein-Uhlenbeck process. This can be found exactly, and can be expressed as a sum of Hermite polynomials and hypergeometric functions, with coefficients chosen to reflect the condition of decay as $v\to \pm  \infty$ \cite{pal2015diffusion, trajanovski2023ornstein}. Alternatively, the Laplace transform can be written in terms of parabolic cylinder functions as described in \cite{veestraeten2015inverse}, and reads
\begin{equation}
    \tilde{\mathcal{P}}_0(v,s| v_0) = \frac{\Gamma(s/\gamma)}{ \sqrt{2 D \pi \gamma}}e^{- \frac{v^2-v_0^2}{4 k_B T}}\mathscr{F}(v,s,v_0)
\end{equation}
where the function $\mathscr{F}$ takes the form
\begin{equation}
    \mathscr{F}(v,s,v_0) =
\left\{
	\begin{array}{ll}
		\mathscr{D}_{-s/\gamma}\left(-\frac{v}{\sqrt{k_BT}}\right)\mathscr{D}_{-s/\gamma}\left(\frac{v_0}{\sqrt{k_BT}}\right)  & \mbox{if } v \leq v_0 \\
\mathscr{D}_{-s/\gamma}\left(\frac{v}{\sqrt{k_BT}}\right)\mathscr{D}_{-s/\gamma}\left(-\frac{v_0}{\sqrt{k_BT}}\right)  & \mbox{if } v \geq v_0 	\end{array}
\right.
\end{equation}
where $\mathscr{D}_{\nu}(z)$ is the parabolic cylinder function of order $\nu$~\cite{NIST:DLMF}. 
The steady-state for the Ornstein-Uhlenbeck process describing velocity with resetting to $v_0$ is thus 
\begin{equation}
    \mathcal{P}(v|v_0) = \frac{\Gamma(r/\gamma)}{ \sqrt{2 k_B T \pi \gamma^2}}e^{- \frac{v^2-v_0^2}{4k_B T}}\mathscr{F}(v,r,v_0)
\end{equation}

\begin{figure}
    \centering
    \includegraphics[width=0.99\linewidth]{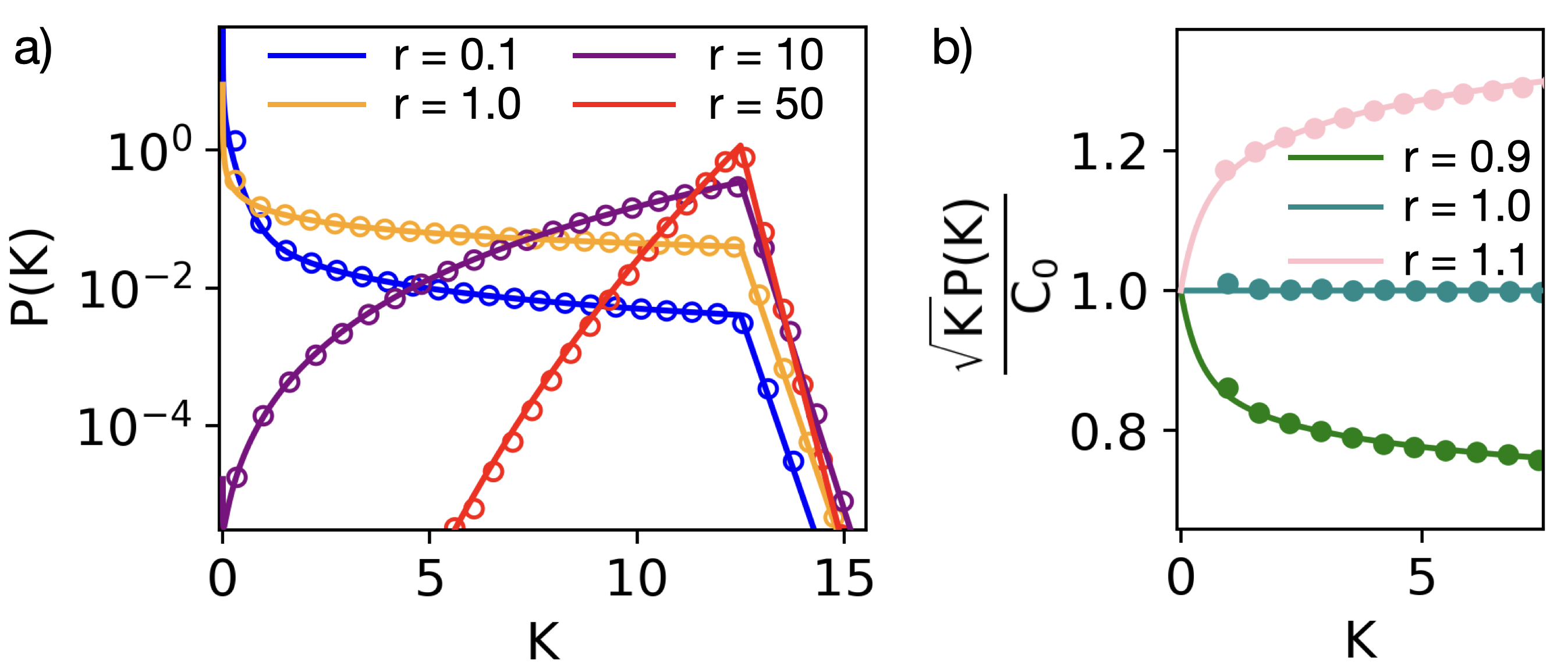}
    \caption{a) Steady state energy distributions for various renewal rates $r$. Dots show numerical simulation data while solid lines are theoretical predictions. For all rates, we see a first-order discontinuity at the resetting energy, common to most resetting problems. A universal $K^{-1/2}$ divergence as $K\to 0$ follows immediately from the quadratic relation between velocity and energy. b) For the regularized density $\sqrt{K} P(K)/C_0$ a shape transition is seen at $r=\gamma$. In all cases, the far tail $K > K_0$ is Boltzmannian with a nonequilibrium power-law correction. Parameters used as $K_R = \frac{1}{2}m v_R^2$, $v_R = 5, m = k_BT = \gamma = 1$.}
    \label{fig:KE_dist}
\end{figure}

For $r =0$, one regains the Maxwell-Boltzmann distribution by using the fact that $\mathscr{D}_0(v) = \exp(-v^2/4)$. Taking into account the possibility of resetting to $\pm v_0$ with equal probability, we arrive at
\begin{equation}\label{eq:ss_vel}
    \mathcal{P}(v) = \frac{\Gamma(r/\gamma)}{ \sqrt{2 k_B T \pi \gamma^2}}e^{- \frac{v^2-v_0^2}{4k_B T}}\frac{\mathscr{F}(v,r,v_0)+\mathscr{F}(v,r,-v_0)}{2}
\end{equation}
Together with Eq. (\ref{eq:ss_ke}), this gives the exact steady state distribution of kinetic energy. Figure~\ref{fig:KE_dist}~(a) shows the energy distribution, showing a kink at the renewal energy, as is expected. In the absence of renewal events, the kinetic energy distribution exhibits a $K^{-1/2}$ divergence as $K \to 0$, a direct consequence of the quadratic relation between velocity and energy, followed by an exponential decay $\sim e^{-\beta K/2}$ at large $K$, consistent with equilibrium dynamics.

With the introduction of stochastic energy renewals, the kinetic energy distribution displays two qualitatively distinct behaviors for $K < K_0$ and $K > K_0$. 

For $K < K_0$, the divergence $P(K) \sim K^{-1/2}$ as $K \to 0$ remains universal, again due to the underlying $v^2 \sim K$ relation. To analyze finer structure in this regime, we consider the regularized form $\sqrt{K}P(K)$. Using the asymptotic expansion of the parabolic cylinder functions $\mathcal{D}_{\nu}(z)$, we find~\cite{NIST:DLMF}, 
\begin{align}
\sqrt{K}P(K) = C_0 + \left(r - \frac{\gamma}{m}\right)C_1 K + O(K^2),
\label{nearorigin}
\end{align}
where $C_0, C_1 > 0$ [see Appendix for exact values]. This indicates a shape transition in the small-energy distribution around the critical point $r = \gamma/m$. This is shown in Fig.~\ref{fig:KE_dist} (b), where we see that the regularized steady state transitions from increasing to decreasing  as the critical resetting rate is crossed.

For $K > K_0$, the distribution decays monotonically. Again using the large-argument asymptotes of $\mathcal{D}_{\nu}(z)$, we find that the tail of the distribution follows,
\begin{align}
P(K) \sim K^{-m r/(2\gamma)} e^{-\beta K/2}.
\end{align}
This power-law prefactor modulates the Boltzmann-like exponential decay and reflects the nonequilibrium nature of the system. As $r \to 0$, the distribution recovers the standard Maxwell–Boltzmann form. Non-Boltzmannian tails of this kind are a characteristic of driven dissipative systems, and have been seen in heated granular gases~\cite{rouyer2000velocity,van2004velocity}.

\section{Spatial dynamics}\label{sec:spacex}

Having fully characterised the energetics of the system—including the emergence of enhanced kinetic energy fluctuations due to random energy renewals, we now turn to the spatial dynamics of the particle. Since the position evolves through the integration of a fluctuating velocity, the nature of these kinetic fluctuations is expected to have a direct impact on the particle's displacement statistics. Moreover, position fluctuations are typically more accessible in experiments, making them a natural observable to probe the consequences of nonequilibrium energy injection. 

In the following, we consider one particular case of energy renewal, where the energy renewals correspond to velocity resetting events defined via Eqs.~\eqref{velocity-reset} and \eqref{reset-dist}. The joint distribution of position and velocity $ \mathcal{P}(x,v,t)$\footnote{We denote the joint-distribution, and the marginals by the same letter for notational simplicity} due to this model of energy renewals is
\begin{align}
    \partial_t \mathcal{P}(x,v,t)&=-\partial_v\left(-\frac{\gamma v}{m}\mathcal{P}(x,v,t)-\bar D\partial_v  \mathcal{P}(x,v,t)\right)\cr
    &-v\partial_x \mathcal{P}(x,v,t)
    -r\mathcal{P}(x,v,t)\cr
    &+\frac{r}{2}\mathcal{P}(x,t)(\delta(v-v_0)+\delta(v+v_0))\label{eq:FP}
\end{align}
where $\bar D=\gamma k_BT/m^2$.
In the following, we explore how the stochastic energy renewals influence the evolution of position fluctuations and shape the spatial signatures of the dynamics.

\subsection{Position fluctuations}
A natural way to characterize the spatial behavior of the particle is by studying its position moments and their coupling to velocity. To this end, we note that, the general cross-correlation $\mathcal M_{k,n}(t)=\langle x^k(t)v^n(t)\rangle$ is obtained by multiplying the Fokker-Planck equation by  $x^k(t)v^n(t)$ and then integrating over $x$ and $v$,
\begin{align}
    &\left[\partial_t+\frac{n\gamma}{m}+r\right]\mathcal M_{k,n}(t)=\bar Dn(n-1)\mathcal M_{k,n-2}(t)\label{general-moments}\\
    &+k \mathcal M_{k-1,n+1}(t)
    +\frac r2\left(\frac{\alpha}{\beta m}\right)^{n/2}\mathcal M_{k,0}(t)\left(1+(-1)^n\right)\nonumber
\end{align}
These can be solved systematically using the initial condition $\mathcal M(0,0,t)=1$, and $\mathcal M(k,n,0)=0$ (we assume $x=v=0$ at $t=0$). The symmetry of the problem ensures that $\mathcal M_{2k+1,2n}(t)=\mathcal M_{2k,2n+1}(t)=0$, i.e., the cross-correlations are non-zero when $k+n$ is even.

Using Eq.~\eqref{general-moments}, the position fluctuations are given by the integral 
\begin{align}
   \mathcal  M_{2,0}(t)=2\int_0^t \mathcal M_{1,1}(t')dt'
\end{align}
where the integrand on the right-hand-side can again be obtained from Eq.~\eqref{general-moments} through
\begin{align}
   \mathcal  M_{1,1}(t)=\int_0^t dt' e^{-(\gamma/m+r)(t-t')}M_{0,2}(t').
\end{align}
Further $\mathcal M_{0,2}(t)$ depends only on the normalization $\mathcal M_{0,0}(t)$, and is given by,
\begin{align}
   \mathcal  M_{0,2}(t)&=(2\bar D+rv_0^2)\int_0^t dt' e^{-(2\gamma/m+r)(t-t')}\cr
    &=\frac{2\bar D+rv_0^2}{2\gamma/m+r}\left[1-e^{-(2\gamma/m+r)t}\right]
\label{eq:msd0}
\end{align}
Evaluating the above integrals, one arrives at the result,
\begin{align}
    \mathcal M_{2,0}(t)&=2( 2\bar D+r v_0^2)\left[\frac{me^{-(\gamma/m+r)t}}{\gamma(\gamma/m+r)^2}-\frac{me^{-(2\gamma/m+r)t}}{\gamma(2\gamma/m+r)^2}
    \right.\cr
&\left.-\frac{\frac{3 \gamma }{m}+2 r}{\left(\gamma/m+r\right)^2 \left(2\gamma/m+r\right)^2}+\frac{t}{\left(\gamma/m+r\right) \left(2\gamma/m+r\right)} \right]
\label{eq:msd}
\end{align}

\begin{figure}
    \centering
   \includegraphics[width=\linewidth]{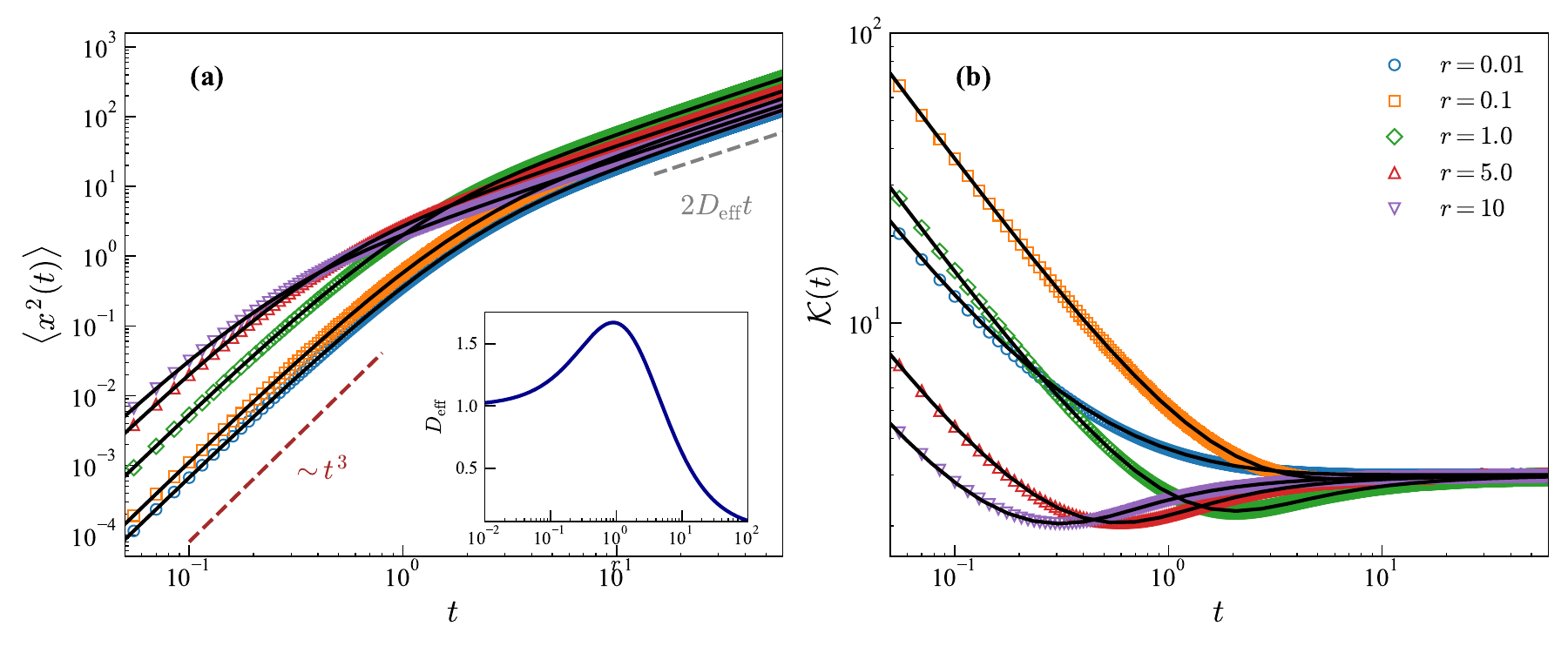}
    \caption{(a) Plot showing the comparison of the mean squared displace of the Brownian particle with energy renewals obtained from numerical simulations (symbols) and theory (solid lines cf. Eq.~\eqref{eq:msd}). The inset shows the variation of the effective diffusion coefficient with respect to the renewal rate $r$ c.f. Eq.~\eqref{deff}. Panel (b) shows a plot of the kurtosis obtained exactly by solving the recursive differential equation and that obtained from numerical simulations. The convergence to unit value at long-times imply the Gaussian behavior at long times. For both the plots $\beta=1,v_0=4,\gamma=m=1$}.
    \label{fig:MSD_kurt}
\end{figure}

At short-times the particle shows a super ballistic growth,
\begin{align}
    \mathcal  M_{2,0}(t)\approx\frac{2 \bar D+r v_0^2}{3}\,\, t^3 
\end{align}
This is compared to the same obtained numerical simulations in Fig.~\ref{fig:MSD_kurt}~(a). This can be understood from the fact that, even in the absence of renewals, starting from $(x(0),v(0))=(0,0)$, the friction $\gamma v$ is small, and the particle behaves like a random acceleration process, leading to the $~t^3$ growth~\cite{wang1945theory,duplat2013superdiffusive}. We see that the introduction of energy renewals leads to a correction to the coefficient to the leading order behavior. At long times, however, the particle still shows a diffusive growth with an effective diffusion coefficient
\begin{align}
D_{\text{eff}}=  \frac{2 (\gamma +\alpha  m r)}{\beta(\gamma +m r) (2 \gamma +m r)}. 
\label{deff}
\end{align}
If the energy renewals are done to the typical energy scale $\alpha=1/2$, then $D_{\text{eff}}=(mr+\gamma)^{-1}$, indicating that an introduction of any finite energy renewal rate decreases the effective diffusion rate. However, when the energy is renewed to a higher value, more specifically to $\alpha>3/2$, then the effective diffusion shows a non-monotonic behavior with respect to the renewal rate. 

It is interesting to note that in the limit $\gamma/m\ll r$,  the diffusion coefficient can be cast in a form similar to that of a one-dimensional run-and-tumble particle (RTP), $D_{\text{eff}}=v_0^2/r$, with the propulsion speed $v_0^2=2\alpha/(\beta m)$. This is suggestive of the fact that in this limit the Brownian particle with energy renewals behaves like an RTP. We will investigate this at the level of the full position distribution in the next subsection.

The recursive differential equation for the moments can be systematically solved to arbitrary orders. We also compute the kurtosis $\mathcal{K}(t)=\mathcal M_{4,0}(t)/[\mathcal M_{2,0}(t)]^2$, which measures the deviation from Gaussianity. The lengthy expression for the kurtosis is not explicitly stated for the sake of simplicity, but is plotted in Fig.~\ref{fig:MSD_kurt} (b). We find that the kurtosis goes to 3 at long times indicating a long-time Gaussian behavior.

\subsection{Run-and-tumble dynamics in the slow relaxation regime}
In this section, we explore the parameter regime where the usual velocity relaxation time of the particle $m/\gamma$ is very large, and as seen in the previous section, the effective diffusion coefficient can be recast in the form of an RTP.
In the limit $\gamma/m \ll r$ both the dissipation and diffusive terms are small, and the dynamics is well approximated by
\begin{align}
    \partial_t \mathcal{P}(x,v,t)&=-v\partial_x \mathcal{P}(x,v,t)
    -r\mathcal{P}(x,v,t)\cr
    &+\frac{r}{2}\mathcal{P}(x,v,t)(\delta(v-v_0)+\delta(v+v_0))
\end{align}
We define the position distribution and spatial currents, respectively, as,
\begin{equation}
    \mathcal{P}(x,t) = \int dv P(x,v,t),\quad
    \mathcal{J}(x,t) = \int dv\,  v P(x,v,t)
\end{equation}
which satisfy the continuity equation
\begin{equation}
    \partial_t \mathcal{P}(x,t) = -\partial_x \mathcal{J}(x,t)
\end{equation}
Further, we have the dynamics of the current as
\begin{equation}
    \partial_t \mathcal{J}(x,t) = -\partial_x \int dv v^2 P(x,v,t) - r \mathcal{J}(x,t)
\end{equation}
To obtain a closed equation for the spatial density, one needs a moment closure for the mean squared velocity, since the dynamics of this in general will be coupled to higher-order moments. To proceed, we note that the slow relaxation approximation implies that the velocity distribution will have narrow peaks near $\pm v_0$, justifying the closure
\begin{equation}
    \partial_t \mathcal{J}(x,t) = -v_0^2 \partial_x \mathcal{P}(x,t) - r \mathcal{J}(x,t)
\end{equation}
Combining these equations by taking a second time derivative of the continuity equation results in 
\begin{equation}
    \partial_t^2 \mathcal{P}(x,t) + r\partial_t \mathcal{P}(x,t) = v_0^2 \partial_x^2 \mathcal{P}(x,t)
\end{equation}
which is exactly the Telegrapher equation for an active run-and-tumble particle in one dimension with tumbling rate $r$ and self-propulsion speed $v_0$ \cite{angelani2015run}. This demonstrates that in the regime where the stochastic energy supply is rapid compared to the inertial relaxation timescale, well-established active particle dynamics is expected to emerge. The model hence interpolates between the passive ($\gamma/m \gg r$) and active ($\gamma/m \ll r$) regimes, with an intermediate regime where recurrent relaxation periods and sudden energy injections compete to give rise to a type of intermittent motion.

\subsection{Fluctuations and response}
 The pumping of energy would lead to a violation of the well-known equilibrium form of the fluctuation dissipation theorem (FDT), which states that the linear response to an external force can be determined by the autocorrelation functions in the absence of it. In the following, we explore how FDT is modified due to the energy renewals.

To obtain the autocorrelations of velocity in the presence of energy renewals  $C_{vv}(t_1,t_2)=\la v(t_1)v(t_2)\ra$, we first note that a energy renewal events erases all memory of the velocity process, and thus the velocities are correlated if and only if there is no energy renewal event in between them. Thus we can write a renewal equation~\cite{majumdar2018spectral} for $C_{vv}(t_1,t_2)$ in terms of that without resetting, denoted by $C^0_{vv}(t_1,t_2)$,
\begin{align}
    C_{vv}(t_1,t_2)&=e^{-rt_1} C^0_{vv}(t_1,t_2)\\
    &+r e^{-r(t_1-t_2)}\int_0^{t_2}\, ds e^{-rs} C^0_{vv}(t_1-t_2+s,s)\nonumber
\end{align}
for $t_1>t_2$. The first term accounts from for the contributions where there has been no energy renewal within $[0,t_1]$, while the second term accounts for the contributions coming from all the realizations where the last energy renewal to $\alpha k_BT=\alpha$ happened at a time interval $s$ before $t_2$. The bare correlations are given by,
\begin{align}
    C^0_{vv}(t_1,t_2)=\frac{2\alpha-1}{\beta m}\,e^{-\frac{\gamma}{m}(t_1+t_2)}+\frac{1}{\beta m}\,e^{-\frac{\gamma}{ m}(t_1-t_2)} 
\end{align}
Using the above, the correlations in the stationary state $\{t_1,t_2\}\gg\{r^{-1},m/\gamma\}$ are given by,
\begin{align}
     C_{vv}(t_1,t_2)&=\frac{2(\alpha r+\gamma/m)}{\beta m(r+2\gamma/m)}e^{-(r+\frac{\gamma}{m})(t_1-t_2)}
     \label{eq:2pt}
\end{align}
Note that, for energy renewals consistent with the equipartition of energy, i.e., $\alpha=1/2$, though the stationary value $C_{vv}(0,0)$ remains $k_BT/m$, the decay is faster than the usual underdamped Brownian motion.

Let us consider an external perturbation $V_{\text{ext}}=xf(t)$. The linear response of the velocity due to the perturbation can be written in terms of the linear response function $\chi_{xv}(t)$, 
\begin{align} \label{eq:resp}
\la v(t) \ra_f=\int_0^t\chi_{xv}(t-t')f(t').
\end{align}
The left-hand side can easily be computed by adding a drift term $-(f/m)\partial_v P(x,v,t)$ to the rhs of the Fokker-Planck equation Eq.~\eqref{eq:FP}. Thereafter, moments can be calculated as before, resulting in an explicit equation for $\la v(t)\ra =\mathcal{M}_{0,1}(t)$, given by
\begin{align}
    \left[\partial_t+\frac{\gamma}{m}+r\right]\mathcal{M}_{0,1}(t)&=\frac{f}{m}.
\end{align}
Solving this and comparing to Eq. (\ref{eq:resp}) leads to the velocity response function,
\begin{align}
\chi_{xv}(t)=\frac{1}{m}e^{-(r+\gamma/m)t}.
\label{responsev}
\end{align}
In equilibrium, the Kubo formula relates the response to the time derivative of the cross-correlation,
\begin{align}
    \chi_{xv}(t)=- \beta\frac{d C_{xv}(t,0)}{dt}.
    \label{fdt}
\end{align}
Using $C_{vv}(t,0)$ from Eq.~\eqref{eq:2pt}, we find that the above relation holds with a modified $\beta_{\text{eff}}$, given by,
\begin{align}
\beta_{\text{eff}}=\frac{\beta(r+2\gamma/m)}{2 \left(\alpha r+\gamma/m\right)}
\end{align}

Though the chosen form of effective temperature restores the Kubo relation—an important hallmark of equilibrium linear response theory—it does not yield a consistent thermal description of the system. This inconsistency is reflected in the expression for the effective diffusion coefficient derived in Eq.~\eqref{deff}, which can be recast as $D_{\text{eff}}=\beta_{\text{eff}}^{-1}(\gamma+mr)^{-1}$, highlighting a deviation from the classical Einstein-Sutherland relation.
Furthermore, the effective inverse temperature $\beta_{\text{eff}}$ does not characterize the stationary distribution, which deviates from the equilibrium Maxwell-Boltzmann form and is fundamentally non-Gaussian. This non-existence of consistent effective temperature descriptions are a hallmark of active systems and have been widely discussed in recent literature\cite{loi2008effective,puglisi2017temperature,caprini2021fluctuation,goerlich2022harvesting,hecht2024define}.

Finally, we check the Harada-Sasa theorem ~\cite{harada2005equality,harada2006energy}, which states that the extent of deviation from the equilibrium Fluctuation-response relation Eq.~\eqref{fdt} gives a measure of the rate of energy dissipation in the system,
\begin{align}
    \dot Q =\gamma\int_{-\infty}^{\infty} \frac{d\omega}{2\pi}\left[ \hat C_{vv}(\omega)-\frac{2}{\beta} \Re[\hat\chi_{xv}(\omega)]\right]
\end{align}
where $\hat\cdot$ denotes the Fourier transform defined as $\hat f(\omega)=\int_{-\infty}^{\infty} dt e^{i\omega t}f(t)$, and $\Re[\hat\chi_{xv}(\omega)]$ denotes the real part of the response function in frequency domain. Note that, the $\beta$ in the above relation is just the equilibrium temperature in the absence of energy renewals. Getting the integrand of the rhs of the above equation from Eq.~\eqref{eq:2pt} and \eqref{responsev}, and performing the integral leads to Eq.~\eqref{eq:pi}.

\subsection{Thermodynamic cost of enhanced diffusion in the non-Poissonian case}\label{ssec:thermo}

So far, we have considered the simplest scenario where the energy renewals occur at a constant rate, corresponding to exponentially distributed waiting times between renewal events. While this assumption is analytically convenient, it does not capture the full range of renewal dynamics that may arise in realistic physical or biological systems, where the time between energy injections can follow more complex statistics. In this section, we focus on how energy renewals with an arbitrary waiting time distribution $\psi(t)$ affect the spatial fluctuations of the particle. In particular, we make connection between the transport properties of the particle and its energetics.

To obtain the spatial probability density function, we integrate over velocity to obtain the spatial probability density
\begin{equation}
    \mathcal{P}(x,t|x_0,v_0) \equiv \int dv \mathcal{P}(x,v,t|x_0,v_0).
\end{equation}
In the non-Poissonian case, it is convenient to use a renewal formalism rather than a Fokker-Planck approach to find the densities. The (first) renewal equation, which is standard for renewal processes \cite{evans2020stochastic}, takes the form 
\begin{align}
    &\mathcal{P}(x,t|x_0,v_0) = \Psi(t)\mathcal{P}_0(x,t|x_0,v_0) + \int_0^t d\tau \psi(\tau) \\
    &\times\int dx' \int d v_R \mathscr{P}_R(v_R) \mathcal{P}_0(x',\tau|x_0,v_0) \mathcal{P}(x,t-\tau|x',v_R)\nonumber
\end{align}term
Assuming spatial homogeneous systems, we set $x_0=0$ without loss of generality. Next, we assume that the initial velocity $v_0$ is distributed according to $\mathscr{P}_R(v)$, and define for ease of notation
\begin{equation}
    \wp (x,t)\equiv \int dv_0 \mathscr{P}_R(v_0) \mathcal{P}(x,t|v_0).
\end{equation}
In Fourier-Laplace space, the renewal equation then reads
\begin{equation}
    \hat{\tilde\wp}(k,s) = \mathscr{L}_s[\Psi(t)\hat \wp_0(k,t)] + \mathscr{L}_s[\psi(t) \wp_0(k,t) ]   \hat{\tilde\wp}(k,s) 
\end{equation}
which can be solved to arrive at \cite{olsen2024dynamics}
\begin{equation}
    \hat{\tilde\wp}(k,s) = \frac{\mathscr{L}_s[\Psi(t)\hat \wp_0(k,t)]}{1-\mathscr{L}_s[\psi(t) \wp_0(k,t) ] } 
\end{equation}
While hard to invert exactly, we can find moments of the spatial density by treating $\hat{\tilde\wp}(k,s)$ as the Laplace-transformed moment generating function. First, we note that the first moment vanishes as long as $\mathscr{P}_R(v)$ is symmetric. The second moment can then be obtained from the above as 
\begin{align}
    \mathscr{L}_s[\langle x^2(t)\rangle_r] &= \frac{\int dv_0 \mathscr{P}_R(v_0)\mathscr{L}_s[\Psi(t) \langle x^2(t)|v_0\rangle_0]}{1-\tilde \psi(s)} \\
    &+  \frac{\tilde \Psi(s)}{[1-\tilde \psi(s)]^2} \int dv_0 \mathscr{P}_R(v_0)\mathscr{L}_s[\psi(t) \langle x^2(t)|v_0\rangle_0]\nonumber
\end{align}
Assuming for a moment that $\psi(\tau)$ has well-behaved moments, we can derive a general expression for the effective diffusivity at late times.  At late times, or small $s$, the second term dominates since $1-\tilde \psi(s) = s \langle \tau \rangle +...$, and we can write 
\begin{equation}
    \mathscr{L}_s[\langle x^2(t)\rangle_r] \approx   \frac{1}{s^2 \langle \tau \rangle} \int_0^\infty dt \psi(t)  \int dv_0 \mathscr{P}_R(v_0) \langle x^2(t)|v_0\rangle_0
\end{equation}
with $\approx$ indicating small $s$ behavior. The effective diffusion coefficient $\langle x^2(t) \rangle_r \approx 2 D_\text{eff}t$ is then obtained, as
\begin{equation}
    D_\text{eff} =   \frac{1}{2 \langle \tau \rangle} \int_0^\infty dt \psi(t)  \int dv_0 \mathscr{P}_R(v_0) \langle x^2(t)|v_0\rangle_0
\end{equation}
This is similar to the results obtained in \cite{olsen2024dynamics,santra2024dynamics}, where intermittent motion was considered.

From this, the effective diffusion coefficient of the particle can be calculated easily by using the mean squared displacement without energy renewals, resulting in  
\begin{align}
    D_\text{eff} &=   \frac{1}{2 \langle \tau \rangle} \bigg( (v_0^2 -\frac{k_BT}{m}) \frac{m^2}{\gamma^2} [1 +\tilde \psi(2\gamma/m) - 2\tilde \psi(\gamma/m)] \nonumber \\
    & + \frac{2k_BT}{\gamma} \langle \tau\rangle + \frac{2k_BTm}{\gamma^2}  [\tilde \psi(\gamma/m)-1] \bigg)
\end{align}
In the limit of slow relaxation $\gamma/m \to 0$ when we expect active motion to emerge, we find the diffusion coefficient $D_\text{eff} = \frac{v_0^2}{2 \langle \tau \rangle}\langle \tau^2\rangle$, consistent with diffusivities of run-and-tumble particles \cite{detcheverry2015non}.

\begin{figure}[t!]
    \centering
    \includegraphics[width=\columnwidth]{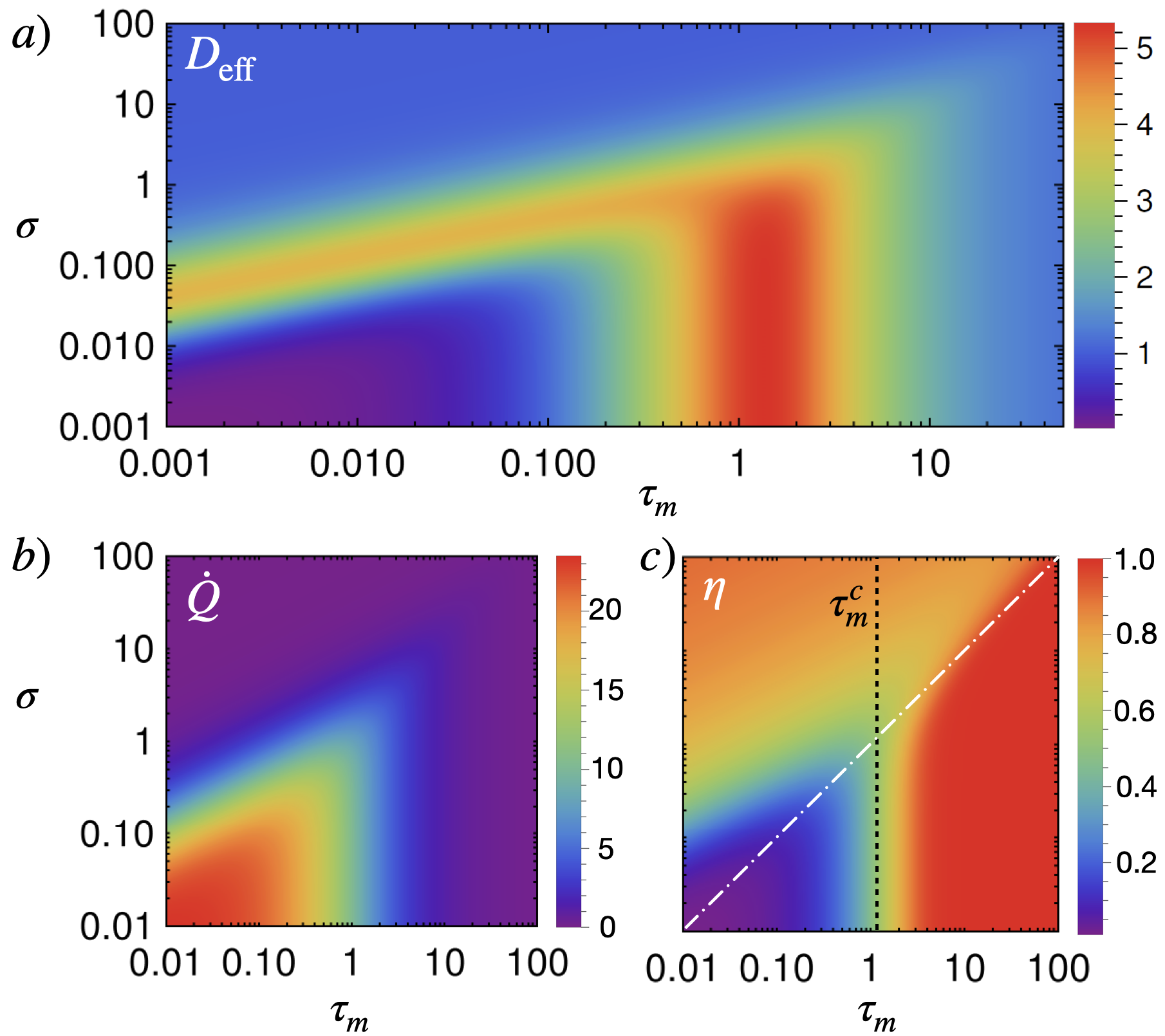}
    \caption{ a) Effective diffusion coefficient as a function of $(\tau_m,\sigma)$. b) Dissipation rate $\dot Q$ as a function of $(\tau_m,\sigma)$. c) Conversion coefficient $\eta$ as a function of $(\tau_m,\sigma)$. Parameters used are $m =\gamma = k_BT = 1$, with $v_0 = 5$ or equivalently $K_0 = 25/2$.}
    \label{fig:deff}
\end{figure}

To get combined insights into the energy expenditure and the spatial dynamics, consider the effective diffusion coefficient at fixed values of $\dot Q$. From earlier discussions about the dissipation rate, we know that at late times $\langle Q\rangle \approx n(K_R - \langle K(\tau)\rangle)$, where $n$ is the number of renewal events in time $t$ and $\langle K(\tau)\rangle$ is averaged both over noise and $\psi(\tau)$. In the non-Poissonian case, assuming well-defined moments of the waiting time density, we have $t = n \langle \tau \rangle$. This leads to the expression
\begin{equation}
    K_0 = \frac{k_B T}{2} +\frac{  \langle \tau \rangle}{1-\tilde \psi(2\gamma / m)} \dot Q
\end{equation}
As a consistency check, we note that in a situation without dissipation, $\dot Q = 0$, we have $K_0 = k_BT/2$, i.e., equipartition of energy. Enforcing a higher dissipation requires a higher amount of energy investment $K_0$ at each renewal. Using this relation to eliminate $K_0$ in the diffusivity gives
\begin{align}
    D_\text{eff} &=  \frac{k_BT}{\gamma} + \frac{k_BTm}{\gamma^2 \langle \tau \rangle} (\tilde \psi(\gamma/m) -1) \\
    &+\frac{m}{\gamma^2}\frac{1-2\tilde \psi(\gamma/m) +\tilde \psi(2\gamma/m)}{ 1-\tilde \psi(2\gamma/m)}\: \dot Q
\end{align}

This result demonstrates a clear relation between thermodynamics and transport; in order to increase the effective diffusion coefficient, more energy must be dissipated. In stochastic thermodynamics, several related ideas where energetics and dynamics are connected has been explored, such as the thermodynamic uncertainty relations that relate entropy and currents, as well as the Harada-Sasa relations discussed in earlier sections. In the current model, a dimensionless coefficient  $\eta[\psi(\tau)] \in [0,1]$ defined as
\begin{equation}
    \eta[\psi(\tau)] \equiv\frac{\gamma^2}{m} \frac{\partial D_\text{eff}}{\partial \dot Q}  =  \frac{1-2\tilde \psi(\gamma/m) +\tilde \psi(2\gamma/m)}{ 1-\tilde \psi(2\gamma/m)}
\end{equation}
couples the effective diffusion coefficient to the dissipation rate, and hence measures how efficiently the energy supply protocol $\psi(\tau)$ and the subsequent energy dissipation is converted into spatial motion. More precisely, if the particle is willing to increase the energy expenditure by an amount $\Delta \dot Q$, the increase in effective diffusion is given by $\Delta D_\text{eff}= \eta[\psi(\tau)] \Delta \dot Q$

Before discussing the general properties of $\eta[\psi(\tau)]$ it is again worth emphasizing a connection to active matter. For a RTP in one dimension \cite{angelani2015run}, the effective diffusion coefficient is $D_\text{RTP} = v_0^2/r$, and the rate of dissipation due to constant-speed motion through a Stokesian medium is $\dot Q = \gamma v_0^2$. Hence $D_\text{RTP} = \dot Q / (\gamma r)$ and the coefficient connecting diffusivity and dissipation rate is simply $1/ (\gamma r)$. When expanding $\tilde \psi(s) = 1 - s \langle \tau \rangle + \frac{1}{2} s^2 \langle \tau ^2 \rangle  +...$ in small inertial relaxation rate compared to renewal rate, one arrives at $\eta[\psi(\tau)] = \langle \tau^2 \rangle/(2 \gamma \langle \tau \rangle ) $ which in the Poissonian case reads $\eta = 1/(\gamma r)$, recovering the result from the standard RTP model.

To get further insights into the conversion of energy into motion, we study the conversion coefficient $\eta[\psi(\tau)]$ for a broader class for waiting times that the exponential (Poissonian) case. For concreteness, we use a gamma distribution with fixed mean $\tau_m$ and standard deviation $\sigma^2$, conveniently parametrized as
\begin{equation}
    \psi(\tau) = \left(\frac{\tau_m \tau}{\sigma^2}\right)^{\frac{\tau_m^2}{\sigma^2}} \frac{\exp(-\tau_m \tau/\sigma^2)}{\tau \Gamma(\tau_m^2/\sigma^2)}
\end{equation}
The Laplace transform is easily found to be $\tilde \psi(s) = (\tau_m/\sigma^2)^{\tau_m^2/\sigma^2} (s + \tau_m/\sigma^2)^{-\tau_m^2/\sigma^2}$, from which the above dynamical quantities can be calculated.

Fig.~\ref{fig:deff} (a) shows the effective diffusion coefficient, displaying non-trivial behaviors both as a function of the mean inter-renewal time and the standard deviations on this time. At sufficiently small fluctuations, an optimal mean time exists, taking a value close to the inertial relaxation timescale $\sim m/\gamma$. For higher values of fluctuation, the diffusion coefficient becomes a monotonically increasing function with $\tau$.  Another aspect worth emphasizing is that while the globally optimal effective diffusion seems to take place close to $\tau_m \approx m/\gamma$ and at small $\sigma$, there also exists regimes where fluctuations are beneficial. For small $\tau_m$, the effective diffusion coefficient can be maximized with respect to $\sigma$, while for large $\tau_m$ no such optimum exists. 

Fig.~\ref{fig:deff} (b) shows the dissipation rate of the process, which has a clear maximum in the regime of rapid deterministic renewals. In this regime the particle  almost always has a higher speed~$\simeq v_0>\sqrt{2/(\beta m)}$ with little fluctuations, leading to a higher rate of dissipation through friction. The conversion coefficient $\eta$ is shown in Fig.~\ref{fig:deff} (c) as a function of the mean inter-renewal time and its standard deviations.  At large $\tau_m$ $\eta$ is high, simply because in this regime renewals are rare, and the particle has time to utilize all the supplied energy before another renewal. However, in this regime the effective diffusion coefficient is also small [see Fig.~\ref{fig:deff} (a)], taking a value close to that in the passive regime. Furthermore, at small $\tau_m$ and $\sigma$ the particle receives rapid deterministic energy supplies, and does not have time to utilize the given energy. Also here the effective diffusion is low. In contrast, close to the regime where the diffusivity is maximized, close to $\tau_m = 1$ (in units of the inertial timescale) $\eta$ also takes an intermediate value. A critical mean inter-renewal time $\tau_m^c = \frac{m}{\gamma} \log(3)$ separates regimes where the conversion coefficient is monotonically increasing with fluctuations $(\tau_m < \tau_m^c)$ and a regime where a non-monotonicity takes place $(\tau_m > \tau_m^c)$, with $\eta$ first decreasing as a function of $\sigma$ before slowly increasing again.

\section{Discussion}\label{sec:concl}

We have introduced a minimal model for a Brownian particle subject to discrete, stochastic energy injections and analyzed its nonequilibrium behavior. By combining standard Langevin dynamics with intermittent renewals to the kinetic energy based on stochastic resetting methods, we were able to explore how random energy renewals affect both the energetics and spatial dynamics of the system. We computed the full kinetic energy distribution and its exact moments, revealing clear signatures of nonequilibrium behavior, such as the breakdown of the equipartition  theorem and shape-transitions in the stationary energy distribution. In addition, we calculated the mean dissipation rate, providing a thermodynamic measure of the system's deviation from equilibrium.

The velocity correlations, while still exponential, have a smaller decay time-scale than in equilibrium. We find that though the associated linear response obeys a modified Kubo relation with an effective inverse temperature $\beta_{\text{eff}}$, restoring an FDT-like structure, this effective temperature fails to characterize either the stationary velocity distribution or the Einstein relation. We also find that the extent of departure from the equilibrium form of the FDT directly characterizes the dissipation in the system through a Harada-Sasa relation.

Extending beyond energetics, we analyzed the spatial dynamics of the particle under both Poissonian and non-Poissonian energy injection protocols. For the Poissonian case, we obtained exact results for the mean squared displacement, and studied the fluctuation dissipation theorem. In the non-Poissonian regime, we calculated a general expression for the effective diffusion coefficient. Crucially, we expressed the diffusivity as a function of the dissipation rate, establishing a direct connection between energetic cost and spatial exploration. This relationship highlights how intermittent, nonthermal driving reshapes both the thermodynamics and transport properties of the system. 

Our work opens up several avenues for future exploration. Active particles in confinement are known to exhibit non-Boltzmannian steady states~\cite{dhar2019run,santra2021direction,caraglio2022analytic}.The intermediate activation regime studied here, where velocity relaxation and renewal timescales are comparable, is expected to display even richer dynamical behavior and provides a promising outlook. The study of first-passage statistics optimized with respect to thermodynamic cost also presents a promising future direction, where trade-offs between energy expenditure and search times could be characterized. Furthermore, in multi-particle systems, imposing a fixed global shared energy budget at each renewal event could give rise to nontrivial collective phenomena even in the absence of direct interactions. It is also natural to generalize the model presented here to higher spatial dimensions. In one dimension, relating energy renewals to a consistent choise of velocity randomization was a natural choice; extending this idea to higher dimensions offers another compelling and more complex avenue of research, where connections to established active matter models like higher dimensional RTPs~\cite{santra2020run} and active Brownian particle models~\cite{basu2018active,santra2021active} can be explored.

\acknowledgements
 K.S.O acknowledges support from the Alexander von Humboldt fellowship program. The authors thank the \emph{Wilhelm and Else Heraeus-Foundation} and organizers of `826. WE-Heraeus-Seminar on Complex Spreading Phenomena: From Bacteria to Innovations', where this work was initiated. 

\appendix
\section{Coefficients in \eqref{nearorigin}}
In this appendix we provide the expressions for the coefficients $C_0$ and $C_1$ that appear in the asymptotic forms of $\sqrt{K}P(K)$ near $k=0$, 
\begin{align}
    C_0&=\frac{\Gamma \left(\frac{r}{2 \gamma }+1\right) \, _1F_1\left(\frac{r}{2 \gamma };\frac{1}{2};\frac{v_0^2 \beta }{2}\right)}{r \sqrt{\frac{1}{\beta  m}} \Gamma \left(\frac{r+\gamma }{2 \gamma }\right)},\\
    C_1&=\frac{ \Gamma \left(\frac{r}{2 \gamma }\right) \, _1F_1\left(\frac{r}{2 \gamma };\frac{1}{2};\frac{\text{v0}^2 \beta }{2}\right)}{2 \left(\frac{1}{\beta }\right)^{3/2} \gamma ^2 \sqrt{m} \Gamma \left(\frac{r+\gamma }{2 \gamma }\right)}.
\end{align}
Here ${}_1F_1(a,b,z)$ denotes the confluent hypergeometric function, that has a convergent series expansion for positive arguments,
\begin{align}
  {}_1F_1(a,b,z)=\displaystyle\sum_{j=1}^{\infty}\frac{(a)_j}{(b)_j}\frac{z^j}{j!}  
\end{align}
where $(a)_j$ represents the Pochhammer symbol. Evidently both $C_0$ and $C_1$ are positive.

\section*{References}
\bibliography{refs.bib}

\end{document}